\documentclass[12pt]{article}
\usepackage{graphicx,amsmath}
\usepackage{color}

\newlength{\dinwidth}
\newlength{\dinmargin}
\begin{document}

\def\be{\begin{equation}}                                                    

\def\lapproxeq{\lower .7ex\hbox{$\;\stackrel{\textstyle                                                    
<}{\sim}\;$}}                                                    
\def\gapproxeq{\lower .7ex\hbox{$\;\stackrel{\textstyle                                                    
>}{\sim}\;$}}                                                    
\def\be{\begin{equation}}                                                    
\def\ee{\end{equation}}                                                    
\def\bea{\begin{eqnarray}}                                                    
\def\eea{\end{eqnarray}}

\begin{flushright}                                                    
IPPP/18/85\\
\today \\                                                    
\end{flushright} 

\vspace{1cm}

\begin{center}
{\Large\bf Black disk radius constrained by unitarity}\\                                                    
                                                   
\vspace*{0.5cm} 

V.A. Khoze$^{a,b}$, A.D. Martin$^a$ and M.G. Ryskin$^{a,b}$\\ 

\vspace*{0.5cm}                                                       
$^a$ Institute for Particle Physics Phenomenology, University of Durham, Durham, DH1 3LE \\                                                   
$^b$ Petersburg Nuclear Physics Institute, NRC Kurchatov Institute, Gatchina, St.~Petersburg, 188300, Russia

 \vspace{1cm}

\begin{abstract} 

We argue that if the elastic proton-proton cross section increases with energy, the Froissart-like high energy behaviour of the elastic amplitude (which corresponds to a `black disk' of radius $R(s)=c\ln s-\beta\ln(\ln s)$) is the only possibility to satisfy the unitarity equation at each value of the impact parameter, $b$. Otherwise the cross section 
of events with Large Rapidity Gaps grows faster than the total cross section at the same $b$. That  is, these `gap' events require {\it maximal} growth of the high-energy (asymptotic) cross section and of the interaction radius $R(s)$ in order to be consistent with unitarity.
\end{abstract}
\end{center}
\vspace{0.5cm}

\section{Introduction}
 It was shown long ago~\cite{GM} that in the so-called `strong coupling' regime (where the cross section increases with energy) the high energy, $\sqrt s$, dependence of the total and elastic cross sections take the form
\be
 \label{sig}
 \sigma_{\rm tot}=C_t(\ln s)^\eta,\quad\;\;\;\;\frac{d\sigma_{\rm el}}{dt}~=~C_{el}(\ln s)^{2\eta} F(B(s)t)\ ,
 \ee 
with the function $F$ chosen to describe the $t$ dependence of the elastic cross section, with the ``slope''
\be\label{slope}
B(s)~=~B_0(\ln s)^\gamma\ ,
\ee
where the parameters $\eta$ and $\gamma$ are limited to the intervals $0\leq \eta\leq \gamma$ and $0\leq\gamma\leq 2$. 

Note that processes with Large Rapidity Gaps (LRG) were not considered specially in ~\cite{GM}.
 In a recent paper~\cite{black} we argued that when we account for events with LRG  the only possibility to satisfy unitarity is to make the  disk completely black. That is, in terms of (\ref{sig},\ref{slope}) to put
$\eta=\gamma$. Moreover, when we consider the contribution of LRG events at the edge of disk (where the disk is not black but 'grey', i.e. partly transparent) we find that the radius of the disk must grow as 
\be
R(s)\propto(\ln s)^{\gamma/2}\propto\ln s.
\ee
 That is the only solution is $\eta=\gamma=2$.

In section 2 we recall the main arguments of ~\cite{black} in favour of black disk asymptotics. In section 3 we study LRG events at large impact parameter $b$. There we show that only in the case of $\gamma=2$ (that is $R(s)\propto \ln s$) does there exist a possibility of screening an increasing LRG cross section in such a way that it does not exceed the total cross section at the same partial wave, that is at the same value of $b$.

\section{Finkelstein-Kajantie  problem }

We first explain the problem. Then we present the solution of the problem and its implications for high-energy proton-proton scattering. Further implications are discussed in Section \ref{sec:3} when we study the behaviour at the edge of the disk.
\subsection{Growth of inelastic cross section with large rapidity gaps}
It was recognized already in the 1960s~\cite{VK-T,FK} that multi-Reggeon reactions, 
\be
pp\to p+X_1+X_2+...+X_n+p,
\ee
where small groups of particles ($X_i$), are separated 
from each other by Large Rapidity Gaps (denoted by $+$ signs), may cause a problem 
with unitarity. Indeed, being summed over $n$ and integrated over the rapidities 
of each group, the cross section of such quasi-diffractive production increases 
faster than a power of $s$. This was  termed 
as the  Finkelstein-Kajantie  disease (FK) in the literature,
see \cite{Abarbanel:1975me} for a review.

Let us explain the problem using the simple example of Central Exclusive 
Production (CEP) of a proton-antiproton pair, as shown in Fig.1.
Since the  proton-proton elastic cross section does not vanish, but increases with energy as $\sigma_{el}\propto (\ln s)^{2\eta-\gamma}$,
 the corresponding contribution to the inelastic cross section reads 
\begin{equation}
\label{t-int}
\sigma^{\rm CEP}~=~N\int_0^Y dy_1\int dt_1 dt_2 ~ 
|A(y_1,t_1)\cdot V\cdot A(Y-y_1,t_2)|^2\propto \int dy_1 \sigma_{\rm el}(y_1)\sigma_{\rm el}(Y-y_1)\ ,
\end{equation}
where the elastic amplitude $A(y,t)$ is normalized in such a way that  $\int dt |A(y,t)|^2=\sigma_{\rm el}(y)$, and the upper rapidity $Y=\ln s/m^2_p$ where $m_p$ is the proton mass. The vertex $V$ describes the central production of a $p\bar p$-pair. In other words we find
\be
\label{cep1}
\sigma^{\rm CEP}~\propto ~(\ln s)^{4\eta-2\gamma+1}.
\ee 
 In the case of a black (or grey) disk of increasing radius when $\eta=\gamma$
 this leads to
 \be
 \sigma^{\rm CEP}~\propto ~(\ln s)^{2\eta+1}~\gg ~\sigma_{\rm tot}~\propto ~(\ln s)^\eta
 \ee  
 \begin{figure}
 \vspace{-6cm}
 \hspace{4cm}
\includegraphics[scale=0.5]{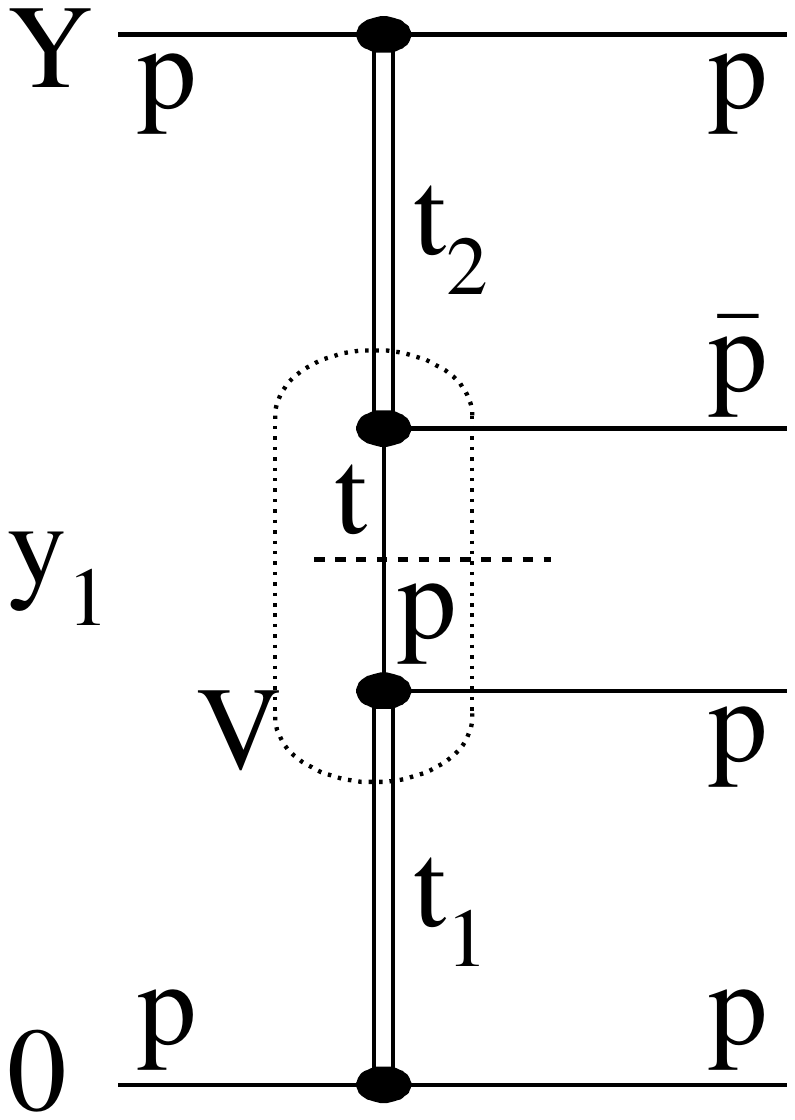}
\vspace{-2cm}
\caption{\sf Central Exclusive Production of a $p\bar p$ pair.}
\label{fig:2}
\end{figure}
The same result can be obtained in impact parameter, $b$, space (see~\cite{black} for details). Moreover
 working in $b$ space we have a stronger constraint since for each value of 
$b$, that is for each partial wave $l=b\sqrt s/2$ of the incoming
 proton pair, we have the unitarity equation
 \begin{equation}
\label{un1}
2\mbox{Im}A(Y,b)=|A(Y,b)|^2+G_{\rm inel}(Y,b)
\end{equation}
and  the `total' cross section, $\sigma(b)_{\rm tot}$ must be less than the 
 corresponding CEP contribution  
 (here $G_{\rm inel}$ denotes the total contribution of all the inelastic channels).
Actually one will face this FK problem in any model where the elastic cross section does 
not decrease with energy. 

At first sight the simplest way to avoid the FK problem is to say that the production 
vertex ($V$ in Fig.1) vanishes.
However this cannot be 
fulfilled. Indeed, as far as we have a non-vanishing high-energy elastic proton-proton cross section we can build diagram Fig.1 
  in such a way that the lower part is just the elastic $pp$-scattering while the upper part corresponds
to the proton-antiproton elastic interaction. Such a diagram is generated by the $t$-channel unitarity equation
\begin{equation}
\mbox{disc}_t~{\bf A}_{12}~=~\sum_j{\bf A}_{1j}^*|j\rangle \langle j|{\bf A}_{j2}\;\; ,
\end{equation}
where in our case $\langle j|$ is the $t$-channel $p$ state.

Note that this contribution 
 is singular at $t=m^2_p$ (where $m_p$ is the proton mass). There are no other
 similar terms corresponding to central exclusive production of $p\bar p$ pair with the same pole singularity. That is, the vertex $V$ contains
  at least one subprocess ($p\bar p$ CEP) which  
cannot be  cancelled identically. See~\cite{black} for more details why this establishes that $V\ne 0$.

\subsection{Black disk solution of the FK problem}
The only known solution of this multi-Reggeon problem comes from `black disk' asymptotics 
of the high-energy cross sections. In such a case the (gap) survival probability, $S^2$, 
of the events with LRGs tends to zero as $s\to\infty$ and the
value of $\sigma^{\rm CEP}$ does not exceed $\sigma_{\rm tot}$. 
\begin{figure} [h]
\vspace{-5cm}
\includegraphics[scale=0.5]{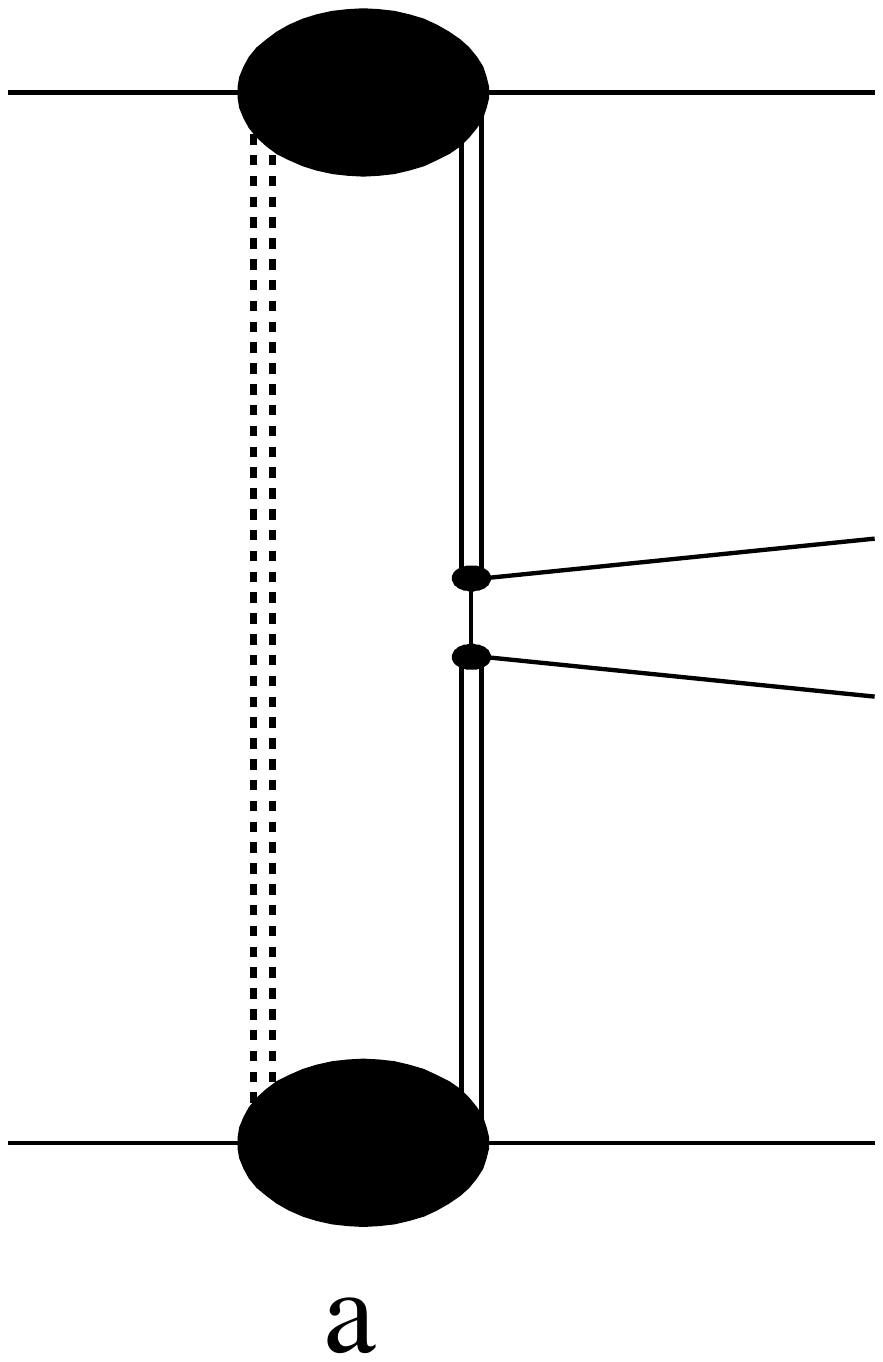}
\hspace{-4cm}
\includegraphics[scale=0.5]{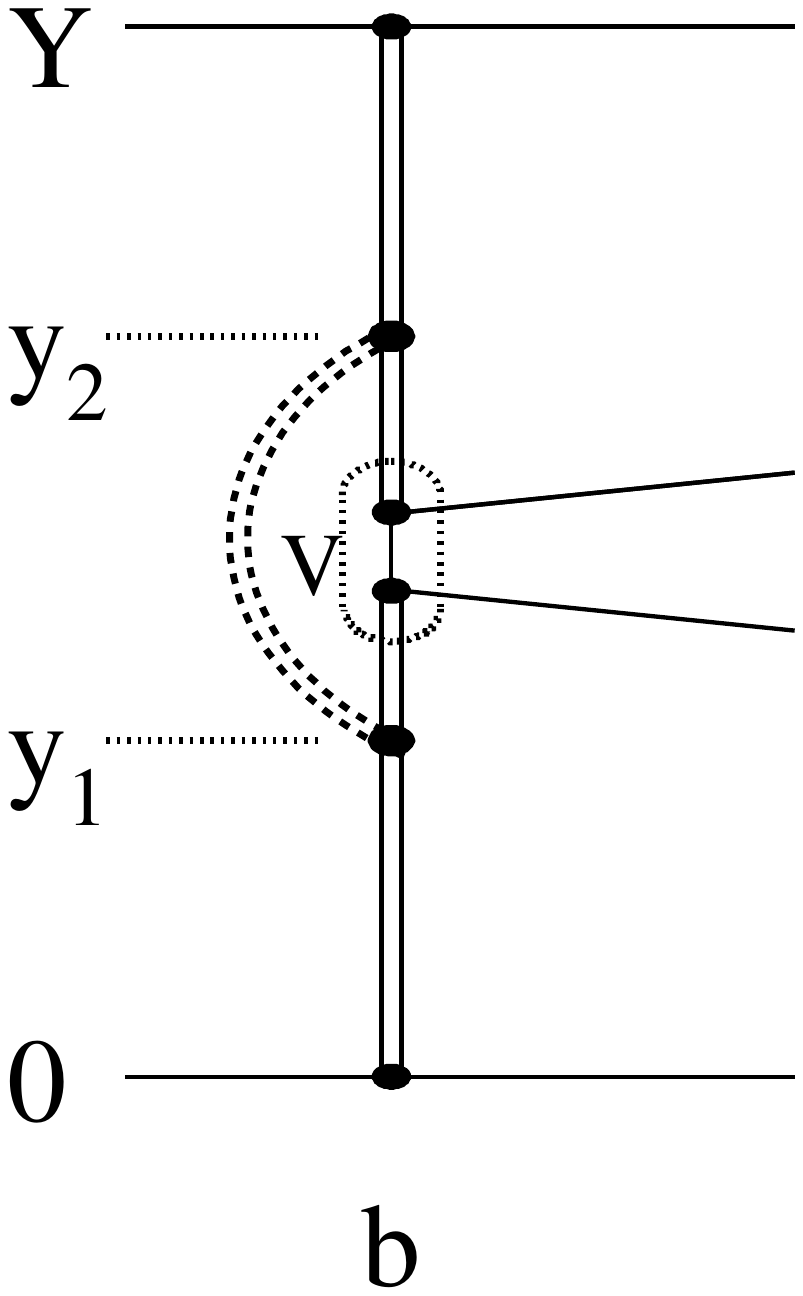}
\vspace{-1.5cm}
\caption{\sf Diagram $a$ shows the amplitude of $p\bar p$ exclusive production screened by an additional inelastic interaction given by the double dotted  line. Diagram $b$ shows the central vertex $V$ screened in some rapidity interval between $y_1$ and $y_2$. }
\label{pp}
\end{figure}
In other words besides the contribution of Fig.1 we have to consider the diagram of 
Fig.2a where the double dotted line denotes an additional (incoming) proton-proton 
interaction. This diagram describes the absorptive correction to the original CEP process 
and has a negative sign with respect to the 
amplitude $A^{(1)}$ of Fig.1. Therefore to calculate the CEP cross section we have 
to square the full amplitude
\begin{equation}
\label{ful}
|A_{\rm full}(b)|^2~=~|A^{(1)}(b)-A^{(2a)}(b)|^2~=~S^2(b)\cdot |A^{(1)}(b)|^2\ ,
\end{equation}
 where the survival factor
 \be
 S^2(b)=|e^{-\Omega(b)}|\ , ~~~~~{\rm with} ~~~~~{\rm Re}\Omega \ge 0 \ ,
 \ee
and $\Omega(b)$ is the opacity of the incoming protons.

 Indeed, in terms of {\bf S}-matrix, the elastic component for a partial wave $l=b\sqrt s/2$ has the form  $S_l=1+iA(b)$, and the unitarity equation (\ref{un1})
reflects the probability conservation condition 
\be
\sum_n{\bf S}^*_l|n\rangle\langle n|{\bf S}_l~=~1~,
\ee
where ${\bf S}_l$ is the component of the ${\bf S}$-matrix corresponding to partial wave $l$. The 
solution of unitarity equation (\ref{un1}) reads
\begin{equation}
\label{el}
A(b)=i(1-e^{-\Omega(b)/2})\ .
\end{equation}
In terms of the partial wave amplitude $a_l$ with orbital moment $l=b\sqrt s/2$ the solution is
\be
a_l~=~i(1-e^{2i\delta_l})~=~i\left(1-\eta_le^{2i{\rm Re}\delta_l}\right)
\ee
where 
\be
\label{eta}
\eta_l=e^{-2{\rm Im}\delta_l}~~{\rm with}~~~ 0\le \eta_l \le 1.
\ee
The above discussion shows that $-\Omega(b)/2$ plays the role of $2i\delta_l$. The elastic component of {\bf S} matrix  
\be
S_l=\exp(2i\delta_l)=\eta_l\exp(2i\mbox{Re}\delta_l).
\ee
 
 The gap survival factor $S^2$ is the probability to observe a pure CEP event where the 
 LRG is not populated by secondaries 
 produced in an additional inelastic interaction shown by the 
 dotted line in Fig.2a. That is according to (\ref{eta})
 \begin{equation}
 \label{gap1}
 |S(b)|^2~=~1-G_{\rm inel}(b)~=~\eta^2=~e^{-{\rm Re}\Omega(b)}\ .
 \end{equation}
 Equation (\ref{gap1}) can be rewritten as (see (\ref{el},\ref{eta}))
 \begin{equation}
 \label{gap2}
 |S(b)|^2=|1+iA(b)|^2=|S_l|^2\ .
 \end{equation}
 
 In the case of {\it black disk} asymptotics\footnote{Recall that the word `black' means the {\em complete} absorption of the incoming state (up to power of $s$ suppressed corrections). That is Re$\Omega(s,b)\to \infty$. `Black disk' means  
 that in some region of impact parameter space, $b < R$, the whole initial 
wave function is absorbed.
  That is, the value of $S(b) = 1 + iA(b) = S_ l \to 0$, i.e. $A(b) \to i$.}
 \be 
 {\rm Re}\Omega(b)\to\infty ~~{\rm and}~~ A(b)\to i,
 \label{eq18}
 \ee
 for $b<R$.   That is, we get $S^2(b)\to 0$. The decrease of the gap survival probability 
 $S^2$ overcompensates the growth of the original CEP cross section (Fig.1), so that 
 finally we have no problem with unitarity.
 
 Recall that this solution of the FK problem was actually realized by Cardy 
\cite{Cardy}, where the reggeon diagrams (generated by Pomerons with intercept 
 $\alpha_P(0)>1$) were considered by {\em assuming analyticity} in the number of Pomerons in a multi-Pomeron  vertex. It was shown that the corresponding absorptive corrections (analogous to that shown in Fig.2a) suppress not only the  growth of  a simplest,
 diagram Fig.1, contribution but the growth of cross sections of  processes with  an arbitrary number of LRGs. 

  Note that at the moment we deal with a one-channel eikonal.  
  In other words in Fig.2 and in the unitarity equation (\ref{un1}) we only account for the pure 
  elastic intermediate states 
  (that is the proton, for the case of $pp$ collisions). In general, there may be 
  $p\to N^*$ excitations shown by the black blobs in Fig.2a.
  The possibility of such  excitations can be included via the  
  Good-Walker~\cite{GW} formalism in terms of G-W eigenstates, $|\phi_i\rangle$,  which 
  diagonalize   the high energy scattering process; that is 
  $\langle\phi_k|A|\phi_i\rangle=A_k\delta_{ki}$. 
  In this case we encounter the FK problem for each state $|\phi_i\rangle $ and we 
  then solve 
  it for the individual eigenstates
  \footnote{For pedagogical purposes it may be heplful to elaborate what would happen if the proton wave function $|p\rangle=\sum_i a_i|\phi_i\rangle$ were to include a `sterile' state $|\phi_s\rangle$ which has zero cross section, that is $A_s=0$. In such a case we would have black disk asymptotic behaviour for all G-W components except that for $|\phi_s\rangle$. However, due to the presence of the $|\phi_s\rangle$ state, then for the whole proton the disk becomes not black but grey. Correspondingly we get a proton elastic cross section $\sigma_{\rm el}<\sigma_{\rm tot}/2$. We emphasize that such a sterile eigenstate must be {\em completely} sterile. It is not sufficient to say that its cross section decreases as a power of the energy. Having a non-zero cross section at low energies, we can consider the diagram Fig.1 in the region of small $y_1$ but very large $Y$. This contribution immediately leads to some non-vanishing amplitude $A_s$ at asymptotically high energies. Thus the presence of a sterile eigenstate appears to be a very extreme hypothesis.  }.

  At first sight it looks sufficient to screen not the whole CEP amplitude, as in Fig.2a, but just the central vertex $V$ as in Fig.2b. 
  Let us consider this so-called enhanced diagram Fig.2b in more detail. Note that we have to integrate over the rapidity-positions $y_1$ and $y_2$ of the `effective' triple-Pomeron vertices. Since the amplitude (shown by the double dotted  line) increases with energy, that is with the size of $|y_2-y_1|$ interval, the main contribution comes from the configurations where $y_1\to 0$ and $y_2\to Y$. In other words the enhanced, Fig.2b, diagram acts as the non-enhanced Fig.2a   graph considered above. 
  
  Moreover, the physical sense of the correction Fig.2b is that simultaneously with an exclusive process some inelastic interaction occurs between the partons placed at $y_1$ and the partons placed at $y_2$.\footnote{The relation between the absorptive correction and the contribution of the processes with a larger multiplicities is given by the AGK cutting rules~\cite{AGK}. The probabilistic interpretation of these AGK rules can be found, for example, in~\cite{LR-AGK}.}
  This interaction violates the `exclusivity' of the process and in this way decreases the cross section of pure CEP events.  If  the central vertex  is screened more or less `locally' (i.e. within a limited $|y_1-y_2|$ interval) then, by cutting the corresponding Pomerons with the help of the AGK rules~\cite{AGK}, we get another LRG process with some more complicated central multiparticle production instead
  of $p\bar p$ production. That is we will get the same FK problem, 
  $\sigma^{\rm CEP}>\sigma_{\rm tot}$ but generated by another group of CEP events.
  
Recall that the inelastic processes generated via the AGK rules by these screening diagrams at any rapidity interval must be included in the whole $G_{\rm inel}$ contribution which describes the correction shown in Fig.2a. That is,  anyway, we get very large probability of {\em inelastic} interactions ($\eta_l\to 0$, i.e. $G_{\rm inel}(b)\to 1$)  and finally arrive in the black disk regime.   
  
 Note also that we cannot avoid the eikonalization which results from the
iterations of two-particle $s$-channel unitarity (\ref{un1}). The opacity $\Omega$ (i.e.
the two-particle irreducible amplitude in (\ref{el})) includes different diagrams but, as
far as $\Omega$ increases with energy ($\Omega\to\infty$ at $s\to\infty$), it
inevitably leads to the black disk regime.

  \section{Edge of the disk \label{sec:3}}

 While the survival factor $S^2$ solves the FK problem for the central part of the black 
 disk, we still have to address the question of what happens at the edge of the disk where 
 the optical density is not large? 
 That is, when $\mbox{Re}\Omega(b)\sim O(1)$. For the large partial waves 
 which occur in this domain we still 
 may have CEP (and other diffractive LRG) cross sections larger than the total cross 
 section corresponding to such $l$-waves.
 
 The solution is provided by the condition  that actually 
 the interaction radius corresponding to a screening amplitude must be larger than the sum of the radii of amplitudes which describe the interactions across
 the gaps (i.e. - the large rapidity intervals). In particular, in the case of Fig.1 the energy/rapidity dependence of interaction radius must satisfy the inequality
 \be\label{ineq}
 R(Y)>R(y_1)+R(Y-y_1)\ .
 \ee 
Using the parametrization given in (\ref{slope}), that is  $R=R_0(\ln s)^{\gamma/2}$, we see that in order to satisfy (\ref{ineq}) we have to choose $\gamma\geq 2$. On the other hand we must satisfy the  Froissart limit $\gamma\leq 2$~\cite{Fr}. That is the only solution is $\gamma=2$; which gives $R\propto \ln s$. 
 
 To be more precise and to provide the {\em inequality} (\ref{ineq})   we have to add the $\ln\ln s$ correction to $R$
 \be\label{Rl}
 R=c\ln s~-~\beta\ln\ln s\ .
 \ee
 Such a correction was obtained for example in~\cite{Az} and ~\cite{RS}. In the latter paper the factor $\ln s$ was replaced by $\ln(s/\sigma_{\rm el})$
 which in the case of $\sigma_{\rm el}\propto \ln^2 s$ generates the $\ln\ln s$ correction in (\ref{Rl}).
 
 Thanks to the fact that (for a large $y_1\sim O(Y)$) the value of $\ln Y=\ln\ln s <\ln y_1+\ln(Y-y_1)$ we get now
 \be\label{Rb}
 R(Y)~>~R(y_1)+R(Y-y_1)\ .
 \ee  
 That is even taking the intermediate amplitudes $A(y_1)$ and $A(Y-y_1)$ at the largest possible impact parameters (at the edge of their disks) we get the total  CEP amplitude, like Fig.1, {\em inside} the completely black disk of the screening amplitude. Thus  the bare  LRG multi-Reggeon contribution will be totally suppressed by the absorptive corrections.

 The mechanism which generates the $\ln\ln s$ correction during the development of the parton cascade (after accounting for processes of diffraction dissociation) was considered in ~\cite{LR}. 
 It was shown in~\cite{LR,DKT} that the same condition (\ref{ineq},\ref{Rl}) provides the 
 possibility to satisfy the $t$-channel unitarity.

\section{Summary}
We emphasize that when high-energy $pp$  cross sections grow with energy, black disk  absorption is the only cure of the FK disease. Thus any asymptotic behaviour of a high energy increasing cross section
 which does not lead to complete absorption is not consistent with multi-particle unitarity. Moreover, in order not to violate the unitarity equation at the edge of disk, where the opacity is not large (Re$\Omega(b)\sim O(1)$), the interaction radius should increase linearly with $\ln s$   
 \be
 R=c\ln s~-~\beta\ln\ln s\ ,
 \label{eq:Rs}\ee
 with a small correction of the order of $\ln\ln s$.
 
 The $R\propto (\ln s)^\delta$ behaviour with $\delta<1$ is rejected since in such a case the cross section of central exclusive events, $\sigma^{\rm CEP}(b)$,
  at large impact parameters $b$ (that is for large partial waves $l=b\sqrt s/2$ occurring at the edge of black disk) grows faster than the total cross section, $\sigma_{\rm tot}(b)$, in the same partial wave. 
 
The remarkable conclusion is that LRG events require {\it maximal} growth of the high-energy (asymptotic) cross section and an interaction radius $R(s)$ of the form of (\ref{eq:Rs})  in order to be consistent with unitarity.

  Finally recall that when we say `black disk' asymptotics we actually refer to the area covered by an `almost' black disk; that is the area where the amplitude $A(b)\simeq i$ is close to black disk limit.   Clearly this area should 
be larger than the area covered by the disk-edge.
 At present in $pp$ scattering at the LHC we are close to black disk
saturation only for $b<b_0=0.2-0.3$ fm while the width of disk-edge is about
1 fm~\cite{Bl}.
 That is the black disk asymptopia that we refer to should start when $b_0$ becomes much larger than 1 fm; say, at $b_0>2~-~3$ fm.
This will correspond to $\sigma_{\rm tot}(pp)=2\pi b^2_0 ~>~ 300~-~ 1000$ mb.

 \section*{Acknowledgements}
 
VAK acknowledges  support from a Royal Society of Edinburgh  Auber award. MGR thanks the IPPP of Durham University for hospitality.


\thebibliography{ }
\bibitem{GM}
V.N. Gribov, Alexander A. Migdal, Zh.Eksp.Teor.Fiz. 55 (1968) 4 ;\\
V.N. Gribov, Alexander A. Migdal 
Sov.Phys.JETP 28 (1969) 784-795.
\bibitem{black}
V.A. Khoze, A.D. Martin, M.G. Ryskin
Phys.Lett. B780 (2018) 352-356,  arXiv:1801.07065 [hep-ph]. 
\bibitem{VK-T} I.A.  Verdiev, O.V. Kancheli, S.G. Matinyan, A.M. Popova
and K.A. Ter-Martirosyan, Sov. Phys. JETP {\bf 19}, 1148 (1964).
\bibitem{FK}
J. Finkelstein, K. Kajantie,  Phys.Lett. {\bf 26B} (1968) 305-307.
\bibitem{Abarbanel:1975me}
   H.~D.~I.~Abarbanel, J.~B.~Bronzan, R.~L.~Sugar and A.~R.~White,
   Phys.\ Rept.\  {\bf 21}, 119 (1975).
\bibitem{Cardy}
  J.~L.~Cardy,
  Nucl.\ Phys.\ B {\bf 75} (1974) 413.

\bibitem{GW} 
   M.~L.~Good and W.~D.~Walker,  
   Phys.\ Rev.\  {\bf 120} (1960) 1857.

\bibitem{AGK}
V.A. Abramovsky, V.N. Gribov, O.V. Kancheli,  Sov.J.Nucl.Phys. 18 (1974) 308-317. 
\bibitem{LR-AGK}
E.M. Levin, M.G. Ryskin, Yad.Fiz. 25 (1977) 849-852. 

\bibitem{Fr}
Marcel Froissart,  Phys.Rev. 123 (1961) 1053-1057.


\bibitem{Az}  	
Ya. Azimov, Phys.Rev. D84 (2011) 056012,
 arXiv:1104.5314 [hep-ph].
 
 \bibitem{RS} 	
V. Singh, S.M. Roy,  Annals Phys. 57 (1970) 461-480

\bibitem{LR} 	
E.M. Levin, M.G. Ryskin, Yad.Fiz. {\bf 27} (1978) 794; Phys. Rept. {\bf 189} (1990) 267 (sect.5).

\bibitem{DKT}  	
M.S. Dubovikov, K.A. Ter-Martirosian,  Zh.Eksperim.I Teor.Fiz. {\bf 73} (1977) 2008; Nucl. Phys. {\bf B124} (1977) 163. 

\bibitem{Bl} 	
Martin M. Block, Loyal Durand, Francis Halzen, Leo Stodolsky , Thomas J. Weiler, Phys. Rev. {\bf D91} (2015) 011501 [arXiv:1409.3196];\\
Dieter Schildknecht, arXiv:1806.05100;\\
A.P. Samokhin,  Phys. Lett. {\bf B786} (2018) 100 [arXiv:1808.07901].
\end{document}